\documentclass[graybox]{svmult}

\usepackage{amsmath}
\usepackage{amsfonts}
\usepackage{amssymb}
\usepackage[english]{babel}
\usepackage{rotating}
\usepackage{booktabs}

\usepackage{mathptmx}       
\usepackage{helvet}         
\usepackage{courier}        
\usepackage{type1cm}        
\usepackage{makeidx}         
\usepackage{graphicx}        
\usepackage{multicol}        
\usepackage[bottom]{footmisc}
\usepackage{multirow}

\DeclareMathOperator*{\argmin}{arg\,min}

\makeindex             

\begin{document}
\title*{Fully reconciled GDP forecasts from Income and Expenditure sides}
\subtitle{\emph{Previsioni riconciliate del PIL dal lato del reddito e della spesa}}
\author{Luisa Bisaglia, Tommaso Di Fonzo and Daniele Girolimetto}
\institute{L. Bisaglia \at Dept. Statistical Sciences, University of Padova, \email{luisa.bisaglia@unipd.it}
	\and T. Di Fonzo \at Dept. Statistical Sciences, University of Padova, \email{tommaso.difonzo@unipd.it}
	\and D. Girolimetto \at Dept. Statistical Sciences, University of Padova, \email{daniele.girolimetto@studenti.unipd.it}}

	%
	%
	\maketitle

\vspace{-1.5cm}

\abstract{
We propose a complete 
reconciliation procedure, resulting in a `one number forecast' of the $GDP$ figure,
coherent with both Income and Expenditure sides' forecasted series,
and evaluate its performance on the Australian quarterly $GDP$ series, as compared to the original proposal by Athanasopoulos {\em et al.} (2019).}
\abstract{\emph{In questo lavoro viene proposta una procedura
di riconciliazione delle previsioni del $PIL$ e delle sue componenti tanto dal lato del Reddito
quanto da quello della Spesa, volta a produrre previsioni coerenti rispetto ad entrambi i lati.
Tale procedura, applicata alle serie trimestrali del $PIL$ australiano, viene posta a confronto con la proposta originale di Athanasopoulos et al. (2019).}}

\vspace{-0.5cm}

\section{Introduction and summary}

\begin{figure}[ht]
\begin{center}
\includegraphics[scale=0.225]{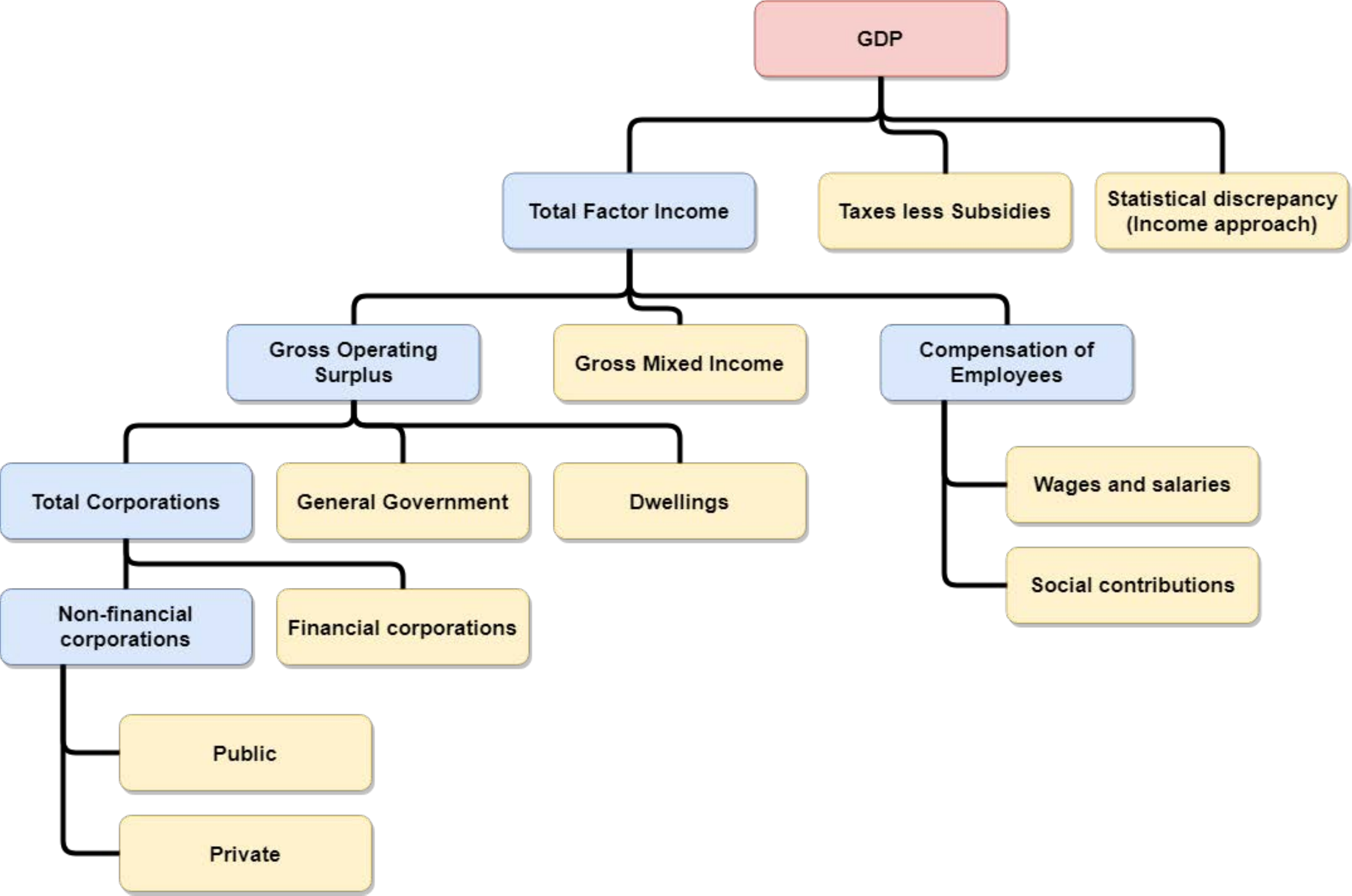}
\caption{Hierarchical structure of the income approach for Australian GDP.
The pink cell contains the most aggregate series. The blue cell contain intermediate-level series and
the yellow cells correspond to the most disaggregate bottom-level series.
Source: Athanasopoulos {\em et al.}, 2019, p. 702.}
\label{fig:AUSINC}
\end{center}
\vspace{-.75cm}
\end{figure}

\vspace{-.25cm}
In a recent paper, Athanasopoulos {\em et al.} (2019, p. 690) propose ``the application of state-of-the-art forecast reconciliation methods
to macroeconomic forecasting'' in order to perform aligned decision making and to improve forecast accuracy.
In their empirical study they consider 95 Australian Quarterly National Accounts time series, describing the
Gross Domestic Product ($GDP$) at current prices from
Income and Expenditure sides, interpreted as two 
distinct
hierarchical structures.
In the former case (Income), $GDP$ is on the top of 15 lower level aggregates (figure \ref{fig:AUSINC}),
while in the latter (Expenditure), $GDP$ is the top level aggregate of a hierarchy of 79 time series (see figures 21.5-21.7 in
Athanasopoulos {\em et al.}, 2019, pp. 703-705).

\begin{figure}[ht]
\begin{center}
\includegraphics[scale=0.45]{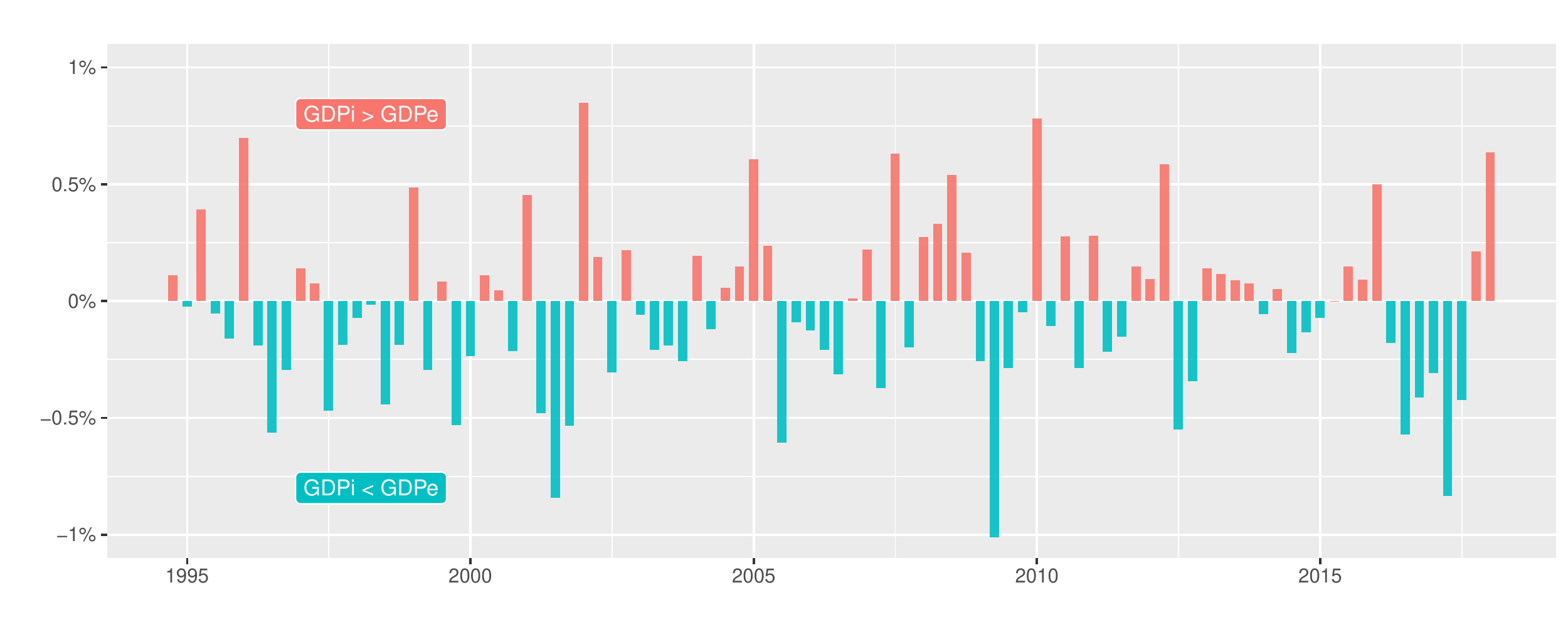}
\caption{Discrepancies in the reconciled 1-step-ahead $GDP$ forecasts from Income and Expenditure sides.
ARIMA base forecasts reconciled according to MinT-shr procedure
(see section \ref{optrec}).
Source data: Athanasopoulos {\em et al.} (2019).}
\label{fig:discrepancy}
\end{center}
\vspace{-.75cm}
\end{figure}

In this paper we re-consider the results of Athanasopoulos {\em et al.} (2019), where
the forecasts of the Australian quarterly $GDP$ aggregates are separately reconciled
from Income ($\widetilde{GDP}^I$) and Expenditure ($\widetilde{GDP}^E$) sides.
This means that $\widetilde{GDP}^I$ and $\widetilde{GDP}^E$ are each coherent within its own pertaining side with the other forecasted values,
but in general $\widetilde{GDP}^I \ne \widetilde{GDP}^E$ at any forecast horizon.
This circumstance could confuse and annoy the user, mostly when the discrepancy is not negligible (see Figure \ref{fig:discrepancy}), 
and calls for a complete reconciliation strategy, able to produce a `one number forecast' of the $GDP$ figure, which is the main target of the paper.

\vspace{-.5cm}

We show that fully reconciled forecasts of $GDP$, coherent with all the
reconciled forecasts from both Expenditure and Income sides, can be obtained through the classical
least squares adjustment procedure proposed by Stone {\em et al.} (1942).
It should be noted that the proposed solution has been considered by van Erven and Cugliari (2015) and Wickramasuriya {\em et al.} (2019)
as an alternative formulation, equivalent to the regression approach by Hyndman {\em et al.} (2011).
As far as we know, however, it has never been applied so far to distinct hierarchies sharing only the top level series.
The procedure can be seen as a forecast combination (Bates and Granger, 1969) - working on different series rather than on the output of multiple models - which makes additional use of external constraints valid for the series and their forecasts.


\vspace{-.75cm}

\section{From single side to complete aggregation constraints}
\vspace{-.25cm}
Denoting with $x_t$ the actual $GDP$ at time $t$, the relationships linking the series of, respectively,
the Income and Expenditure sides hierarchies can be expressed as
\begin{equation}
\label{eq:ySb}
{\bf y}_t^I = {\bf S}^I {\bf b}_t^I, \qquad {\bf y}_t^E = {\bf S}^E {\bf b}_t^E, \quad t=1,\ldots,T, 
\end{equation}
where ${\bf y}_t^I = \left[ x_t \quad {{\bf a}_t^{I}}' \quad {{\bf b}_t^{I}}' \right]'$,
${\bf y}_t^E = \left[ x_t \quad {{\bf a}_t^{E}}' \quad {{\bf b}_t^{E}}' \right]'$,
${\bf b}_t^{I}$ and ${\bf b}_t^{E}$ are $(10 \times 1)$ and $(53 \times 1)$, respectively, vectors of bottom level (disaggregated) series, 
${\bf a}_t^{I}$ and ${\bf a}_t^{E}$ are $(5 \times 1)$ and $(26 \times 1)$, respectively, vectors of higher levels (aggregated) series, 
and
\[
{\bf S}^I = \left[\begin{array}{c} {\bf 1}_{10}' \\ {\bf C}^I \\ {\bf I}_{10}\end{array}\right], \qquad 
{\bf S}^E = \left[\begin{array}{c} {\bf 1}_{53}' \\ {\bf C}^E \\ {\bf I}_{53}\end{array}\right]
\]
are contemporaneous (cross-sectional) summing matrix mapping the bottom level series into the higher-levels variables in each hierarchy, where
${\bf 1}_{k}$ denotes a $(k \times 1)$ vector of ones, ${\bf I}_{k}$ denotes the identity matrix of order $k$, and ${\bf C}^I$ and ${\bf C}^E$
are the $(5 \times 10)$ and $(26 \times 53)$, respectively, matrices of 0's and 1's describing the aggregation relationships between the bottom
level series and the higher level series (apart $GDP$) for Income (${\bf C}^I$) and Expenditure (${\bf C}^E$) sides.
The relationships (\ref{eq:ySb}) can be equivalently written as
\begin{equation}
\label{eq:Uty0side}
{{\bf U}^I}'{\bf y}_t^{I} = {\bf 0}, \qquad {{\bf U}^E}'{\bf y}_t^{E} = {\bf 0}, \quad t=1,\ldots,T, 
\end{equation}
where
${\bf U}^I = \left[\begin{array}{cc} \multicolumn{2}{c}{{\bf I}_6} \\ -{\bf 1}_{10} & -{{\bf C}^I}' \end{array}\right]$,
and
${\bf U}^E = \left[\begin{array}{cc} \multicolumn{2}{c}{{\bf I}_{27}} \\ -{\bf 1}_{53} & -{{\bf C}^E}' \end{array}\right]$
are $(16 \times 6)$ and $(80 \times 27)$ matrices, respectively.
The only variable subject to linear constraints on both the Income and Expenditure sides in expressions
(\ref{eq:ySb}) and (\ref{eq:Uty0side}) being $x_t$ (i.e., $GDP$),
we can express the aggregation relationships linking the 95 `unique' variables as
\begin{equation}
\label{eq:Uty0}
{\bf U}'{\bf y}_t = {\bf 0},  \quad t=1,\ldots,T, 
\end{equation}
where ${\bf y}_t = \left[ x_t \quad {{\bf a}_t^{I}}' \quad {{\bf b}_t^{I}}' \quad {{\bf a}_t^{E}}' \quad {{\bf b}_t^{E}}'\right]'$
is a $(95 \times 1)$ vector ,
${\bf 0}$ is a $(33 \times 1)$ null vector, and ${\bf U}'$
is the following $(33 \times 95)$ matrix:
\begin{equation}
\label{eq:Ut}
{\bf U}' = \left[\begin{array}{ccccc}
                          1 & {\bf 0}_5'             & -{\bf 1}_{10}'         & {\bf 0}_{26}'         &  {\bf 0}_{53}'         \\
                          1 & {\bf 0}_5'             &  {\bf 0}_{10}'         & {\bf 0}_{26}'         & -{\bf 1}_{53}'         \\
                  {\bf 0}_5 & {\bf I}_5              & -{\bf C}^I             & {\bf 0}_{5 \times 26} &  {\bf 0}_{5 \times 53} \\
               {\bf 0}_{26} & {\bf 0}_{26 \times 5}  & {\bf 0}_{26 \times 10} & {\bf I}_{26}          &  -{\bf C}^E
                 \end{array}\right] .
\end{equation}

\vspace{-.75cm}

\section{Optimal point forecast reconciliation}
\label{optrec}
\vspace{-.25cm}
Forecast reconciliation is a post-forecasting process aimed at improving the quality of the {\em base} forecasts
for a system of hierarchical/grouped, and more generally linearly constrained, time series
(Hyndman {\em et al.}, 2011, Panagiotelis {\em et al.}, 2019) by exploiting the constraints that the series in the system
must fulfill, whereas in general the base forecasts don't.
In this framework, as base forecasts we mean the $(n \times 1)$ vector $\hat{\bf y}_{T+h} \equiv \hat{\bf y}_h$
of unbiased point forecasts, with forecast horizon $h > 0$, for the $n>1$ variables of the system.

Following Stone {\em et al.} (1942), we consider the classical measurement model
\begin{equation}
\label{eq:measmod}
\hat{\bf y}_h = {\bf y}_h + {\bf \varepsilon}_h, \quad E\left({\bf \varepsilon}_h\right) = {\bf 0}, \quad
           E\left({\bf \varepsilon}_h {\bf \varepsilon}_h' \right) = {\bf W}_h,
\end{equation}
where $\hat{\bf y}_h$ is the available measurement, ${\bf y}_h$ is the target forecast vector, and
${\bf \varepsilon}_h$ is a zero-mean measurement error, with covariance ${\bf W}_h$, which is a $(n \times n)$ p.d. matrix,
for the moment assumed known. Given a $(n \times K)$ matrix of constant values ${\bf U}$, summarizing the $K$ linear
constraints valid for the $n$ series of the system ($n > K$), in general it is ${\bf U}'\hat{\bf y}_h \ne {\bf 0}$,
and we look for reconciled forecasts $\tilde{\bf y}_h$ such that ${\bf U}'\tilde{\bf y}_h = {\bf 0}$.

The reconciled forecasts $\tilde{\bf y}_h$ can be found as the solution to the linearly constrained quadratic minimization problem:
\[
\tilde{\bf y}_h = \argmin_{{\bf y_h}} \left(\hat{\bf y}_h - {\bf y}_h \right)' {\bf W}_h^{-1} \left(\hat{\bf y}_h - {\bf y}_h\right), \quad
                  \text{s.t. } {\bf U}'{\bf y}_h = {\bf 0},
\]
which is given by
\begin{equation}
\label{eq:ytilde}
\tilde{\bf y}_h = \left[{\bf I}_n - {\bf W}_h{\bf U}\left({\bf U}'{\bf W}_h{\bf U}\right)^{-1}{\bf U}'\right]\hat{\bf y}_h .
\end{equation}
The key item in expression (\ref{eq:ytilde}) is matrix ${\bf W}_h$, which is generally unknown and must be
either assumed known or estimated. In agreement with Athanasopoulos {\em et al.} (2019), 
denoting with
$ \widehat{\bf W}_1$
the $(n \times n)$ covariance matrix of the
in-sample one-step-ahead base forecasts errors of the $n$ series in the system,
we consider 3 cases:
\begin{itemize}
\item OLS: ${\bf W}_h = \sigma^2 {\bf I}_n$
\item WLS: ${\bf W}_h = \widehat{\bf W}_D = \text{diag}\{\hat{w}_{11}, \ldots, \hat{w}_{nn}\}$
\item MinT-shr: ${\bf W}_h = \widehat{\bf W}_{shr} = \lambda\widehat{\bf W}_D + (1 - \lambda)\widehat{\bf W}_1 $
\end{itemize}
where
$\widehat{\bf W}_{shr}$ is the shrinked version of $\widehat{\bf W}_1$, with diagonal target and shrinkage intensity parameter $\lambda$ proposed by
Sch{\"a}fer and Strimmer (2005) (more details can be found in Wickramasuriya {\em et al.}, 2019).

\vspace{-.75cm}

\section{The accuracy of the reconciled forecasts of the Australian GDP} 
\vspace{-.35cm}
According to the notation of the previous section, for the complete Australian $GDP$ accounts from both Income and Expenditure sides,
it is $n=95$, $K=33$, and matrix ${\bf U}'$ is given by (\ref{eq:Ut}). In addition, the available time series span over the period
1984:Q1 - 2018:Q4.

Base forecasts for the $n=95$ separate time series have been obtained by Athanasopoulos {\em et al.} (2019) through simple univariate
ARIMA models\footnote{The R scripts, the data and the results of the paper by Athanasopoulos {\em et al.} (2019)
are available in the github repository located at \url{https://github.com/PuwasalaG/Hierarchical-Book-Chapter}.},
selected using the {\tt auto.arima} function of the R-package {\tt forecast}.
We did not change this first, crucial step in the forecast reconciliation workflow,
since 
the focus is on the potential of forecast reconciliation.\footnote{Athanasopoulos {\em et al.} (2019) point out that this fast and flexible approach performs
well in forecasting Australian GDP aggregates, even compared to other more complex methods.}

Our reconciliation proposal is applied within the same forecasting experiment designed by Athanasopoulos {\em et al.} (2019).
They consider forecasts from $h = 1$ quarter ahead up to $h = 4$ quarters ahead using an {\em expanding} window, where
the first training sample is set from 1984:Q4 to 1994:Q3 and forecasts are produced for 1994:Q4 to 1995:Q3.
The base forecasts are reconciled using OLS, WLS and MinT-shr procedures, and the accuracy is measured by the Mean Squared Error ($MSE$).

\begin{figure}[ht]
\begin{center}
\includegraphics[scale=0.55]{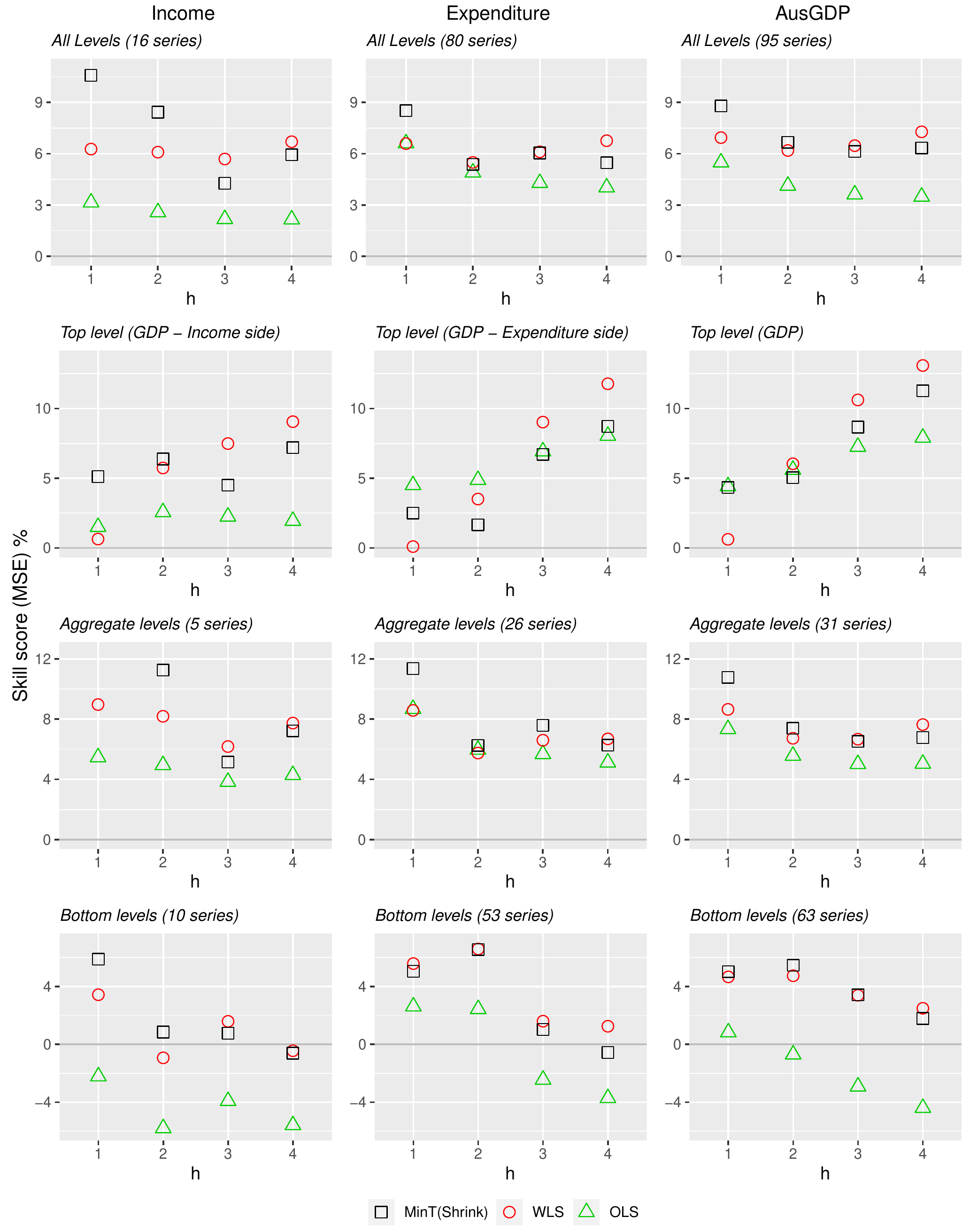}
\caption{Skill scores for reconciled point forecasts from alternative methods
(with reference to base forecasts) using MSE.}
\label{fig:skillscores}
\end{center}
\vspace{-.75cm}
\end{figure}

Figure \ref{fig:skillscores} shows the {\em skill scores} using $MSE$, that is the percentage changes in $MSE$ registered by
each reconciliation procedure, relative to base forecasts, computed such that positive values
signal an improvement in forecasting accuracy over the base forecasts.
The left and the central columns of the figure refer to the results
for the Income and Expenditure sides variables separately considered, while
the right column shows the results of the procedure proposed in this paper.


The results confirm also for the enlarged system the findings of Athanasopoulos {\em et al.} (2019, p. 709):
\vspace{-.25cm}
\begin{itemize}
\item reconciliation methods improve forecast accuracy relative to base forecasts;
\item negative skill scores are registered only for OLS-reconciled forecasts of bottom level series ($h=2,3,4$);
\item MinT-shr is the best reconciliation procedure in most cases.
\end{itemize}
\vspace{-.25cm}
In addition, looking at the second row of figure \ref{fig:skillscores}, we see that for any forecast horizon
the improvements in the unique $GDP$ reconciled forecasts are always larger than those registered
for $\widetilde{GDP}^E$. The same happens with $\widetilde{GDP}^I$, $h=3,4$, while for $h=1,2$ the skill scores are very close.

\vspace{-0.6cm}


\end{document}